\documentclass[aps,pre,showpacs,showkeys,twocolumn,groupedaddress,floatfix]{revtex4-1}
\usepackage{graphicx}
\usepackage{hyperref}
\usepackage{amsthm,amsmath,amssymb}  % AMS

\begin{document}

\title {Impact of defects on percolation in random sequential adsorption of linear $k$-mers on square lattice}

\author{Yuri Yu. Tarasevich}
\email[Correspondence author: ]{tarasevich@aspu.ru}
\affiliation{Astrakhan State University, Astrakhan, Russia}
\author{Valeri V. Laptev}
\affiliation{Astrakhan State University, Astrakhan, Russia}
\affiliation{Astrakhan State Technical University, Astrakhan, Russia}
\author{Nikolai V. Vygornitskii}
\author{Nikolai I. Lebovka}
\email{lebovka@gmail.com}
\affiliation{Institute of Biocolloidal Chemistry named after F.D. Ovcharenko, NAS of Ukraine, Kiev, Ukraine}

\date{\today}

\begin{abstract}
The effect of defects on the percolation of linear $k$-mers (particles occupying $k$ adjacent sites) on a square lattice is studied by means of Monte Carlo simulation. The $k$-mers are deposited using a random sequential adsorption mechanism. Two models, $L_d$ and $K_d$, are analyzed. In the $L_d$ model, it is assumed that the initial square lattice is non-ideal and some fraction of sites, $d$, is occupied by non-conducting point defects (impurities). In the $K_d$ model, the initial square lattice is perfect. However, it is assumed that some fraction of the sites in the $k$-mers, $d$, consists of defects, i.e., are non-conducting. The length of the $k$-mers, $k$, varies from 2 to 256.  Periodic boundary conditions are applied to the square lattice. The dependencies of the percolation threshold concentration of the conducting sites, $p_c$, vs the concentration of defects, $d$, were analyzed for different values of $k$. Above some critical concentration of defects, $d_m$, percolation is blocked in both models, even at the jamming concentration of $k$-mers. For long $k$-mers, the values of $d_m$ are well fitted by the functions $d_m \propto k_m^{-\alpha}-k^{-\alpha}$ ($\alpha = 1.28 \pm 0.01$, $k_m = 5900 \pm 500$) and $d_m \propto \log (k_m/k)$ ($k_m = 4700 \pm 1000$ ), for the $L_d$ and $K_d$ models, respectively. Thus, our estimation indicates that the percolation of $k$-mers on a square lattice is impossible even for a lattice without any defects if $k\gtrapprox 6 \times 10^3$.
\end{abstract}



\maketitle

\section{Introduction}
\begin{figure*}
  \centering
  \includegraphics[clip=on,width=\textwidth]{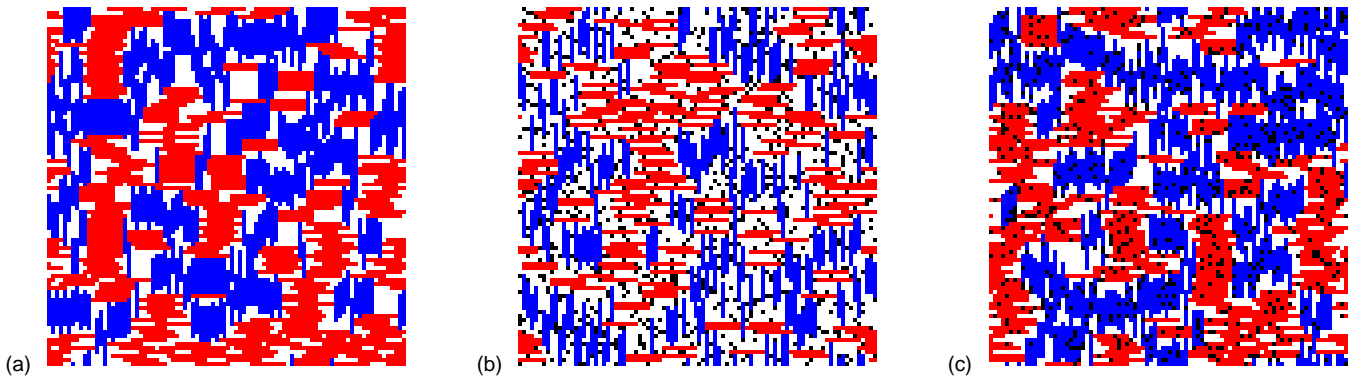} \\
  \caption{Jamming states: (a) a perfect lattice filled with defect-free linear $k$-mers, (b) a lattice with defects filled with defect-free $k$-mers, (c) a perfect lattice filled with defective $k$-mers. Online: Horizontal $k$-mers are shown in red, vertical $k$-mers are shown in blue, empty sites are shown in white, defects are shown in black. Print: gray-scale. $k=9$, fragment of a lattice $90 \times 90$ sites.\label{fig:lattices}}
\end{figure*}

The adsorption of large particles such as colloids, proteins or nanotubes onto substrates
can be treated and studied as random sequential adsorption (RSA)~\cite{Evans1993RMP}.
In RSA, objects randomly deposit onto the substrate; this process is irreversible, and the newly placed objects cannot overlap or pass through the previously deposited ones (for details see, e.g. a review by Evans~\cite{Evans1993RMP}). Very often,  irreversible RSA, without detachment or diffusion is considered.
If the concentration of the deposited objects is large enough they form a path from one side of the system to its opposite side. Below this concentration, a spanning path does not exist; while above it, there exists a spanning component  of the system size order. This concentration is denoted as the percolation threshold~\cite{Stauffer}.
By placing objects at random onto a surface, but adsorbing only those that do not overlap previously adsorbed objects, one will finally reach a jamming limit beyond which no more objects can be adsorbed~\cite{Evans1993RMP} (see an example of a jamming state in Figure~\ref{fig:lattices}a).

One of the simplest examples of a substrate is a square lattice. A simple but very attractive instance of the type of object being adsorbed  is a stiff rod, also denoted as stiff-chain or stick~\cite{Becklehimer1992}, line segment~\cite{Leroyer1994PRB}, rigid rod, needle~\cite{Vandewalle2000epjb}, linear $k$-mer, etc. Here $k$ means the length of the object. For uniformity,  hereinafter we will use the term $k$-mer.

Many important findings regarding percolation and jamming of completely disordered $k$-mers~\cite{Becklehimer1992,Leroyer1994PRB,Bonnier1994,Vandewalle2000epjb,Kondrat2001PRE,Cornette2003epjb} or partially aligned $k$-mers~\cite{Longone2012PRE,Lebovka2011PRE,Tarasevich2012PRE,Romiszowski2013} have been reported over the last two decades.

Becklehimer and Pandey found~\cite{Becklehimer1992} that the percolation threshold decreases if the length of the sticks increases as
$$
p_c(k) \sim k^{-1/2},
$$
where $k$ is the stiff-chain (stick) length. $k$ varies up to 20. The jamming
coverage decreases with the chain-length.

Leroyer and Pommiers investigated the percolation of line segments (linear $k$-mers with $k$ values up to 40) and  found that the percolation threshold initially decreases and then increases as the length of the segments  increases~\cite{Leroyer1994PRB}.

Bonnier et al. studied the deposition of line segments with $k$ values up to 512 on a two-dimensional square lattice and found that the jamming concentration goes down asymptotically to $0.660 \pm 0.002$ as the length of the rods increases~\cite{Bonnier1994}.

Vandewalle et al. explored the random sequential deposition of needles (linear $k$-mers with $k$ values up to 10) and suggested that the percolation threshold and the jamming concentration decrease as the length of the rods increases
$$
p = C \left( 1 - \gamma \left(\frac{ k - 1}{k} \right)^2 \right)
$$
both for percolation and for jamming. Here $k$ is the length of the needles (sticks, rods, linear $k$-mers), while $C$ and $\gamma$ are the fitting parameters~\cite{Vandewalle2000epjb}.

Kondrat and P\c{e}kalski (Ref.~\cite{Kondrat2001PRE}) examined the percolation and jamming of linear segments on a square lattice ($k$ up to 2000) and found that the percolation threshold, $p_c$, and the jamming concentration, $p_j$, initially decrease but then increase as the length of the rods increases
$$
p_c /p_j \sim 0.50 + 0.13 \log_{10} k.
$$
One can easily calculate $p_c = p_j$ if $k \approx 0.7 \times 10^4$.

Cornette et al. studied the percolation of linear segments of size $k$ and $k$-mers of different structures ($k$ varies up to 15) on a square lattice and reported that the percolation threshold exponentially decreases with the length of the rods~\cite{Cornette2003epjb}. The percolation threshold as a function of $k$ for linear segments is fitted by
$$
p_c(k) = p^*_c + \Omega \exp \left( - \frac{k}{\kappa} \right),
$$ where the fitting parameters are
$p^*_c = 0.461 \pm 0.001$, $\Omega = 0.197 \pm 0.02$, $\kappa = 2.775 \pm 0.02$.

Longone et al.~\cite{Longone2012PRE} studied the deposition of aligned rigid rods of length $k$ up to 12 and  found that the percolation threshold for completely  ordered deposition decreases if the length of the rods increases, which is similar to the case with disordered deposition~\cite{Cornette2003epjb}.

Recently, jamming~\cite{Lebovka2011PRE} and percolation~\cite{Tarasevich2012PRE} of linear $k$-mers have been intensively studied for values of $k$ up to 512  for percolation and up to 256 for jamming. It was demonstrated that the jamming concentration continuously decreases as the length of the $k$-mer  increases. On the other hand, the percolation threshold initially decreases and then increases with increasing $k$. For completely disordered systems, a conjecture has been offered that percolation is impossible if $k$ exceeds approximately  $1.2 \times 10^4$~\cite{Tarasevich2012PRE}. Direct verification of the conjecture is very time-consuming and problematic even with a high-performance computer. Less laborious indirect methods for testing the hypothesis would be very appealing. We should emphasize that the estimation in~\cite{Tarasevich2012PRE} has the same order as the estimation in~\cite{Kondrat2001PRE}, nevertheless the results for large values of $k$ have rather large errors in both works.

However, the deposition of particles without defects onto a completely regular lattice would seem to be too idealized a model. In the real world, a substrate may be not perfect, i.e. initially, some impurities may be already present on the substrate. In most practical cases, the surfaces are chemically heterogeneous and contain defects~\cite{Adamson1997}. In theoretical works, a lattice with impurities is sometimes treated as a diluted lattice~\cite{Cornette2006PLA}. Such a lattice is built by randomly selecting a fraction of the sites which are then considered forbidden for the succeeding deposition of any objects. A jamming state on a diluted square lattice is presented in Figure~\ref{fig:lattices}b. For uniformity within the rest of the text, we shall use the term 'defect' here and below.

The structure of elongated particles, e.g., carbon nanotubes, may also be highly heterogeneous due to their chemical functionalization; moreover, a length polydispersity may also be present~\cite{Wepasnick2010}.

In practice, both kinds of imperfection may occur, i.e. one has to consider the deposition of objects
with defects onto a substrate with defects.

That is why the study of defective systems involving $k$-mers is very important and many previous works  have been related to these problems.

The jamming and percolation of $k$-mers on disordered (or heterogeneous) substrates with defects (or impurities) has attracted great attention~\cite{Ben-Naim1994JPhysA,Lee1996JPhysA,Budinski-Petkovic2002,Kondrat2005,Kondrat2006, Cornette2006JCP,Cornette2006PLA,Cornette2011PhysA,Budinski-Petkovic2011,Budinski-Petkovic2012}. A lattice with defects is built by randomly selecting a fraction of insulating monomers~\cite{Cornette2006JCP,Cornette2006PLA,Budinski-Petkovic2012}
or $k$-mers~\cite{Kondrat2005,Kondrat2006} which are considered forbidden for the succeeding deposition of any objects. The jamming limits $p_j$ for the deposition of $k$-mers onto a one-dimensional line and a two-dimensional disordered square lattice were calculated using the Monte Carlo method~\cite{Lee1996JPhysA}.
Note that the total jamming coverage (i.e., the value of $p_j+d$) decreased when the concentration of impurities $d$ increased and reached a minimum which depended on $k$~\cite{Lee1996JPhysA, Budinski-Petkovic2002}. For the problem of $k$-mer deposition onto a square lattice, it has been established that, upon increasing $d$, the percolation threshold $p_c$ grows up to a maximum value $p_c^m(d_m)$, and that there is no percolation above $d_m$~\cite{Cornette2006JCP,Cornette2006PLA,Cornette2011PhysA}.

Cornette et al.~\cite{Cornette2006JCP,Cornette2006PLA} investigated, numerically, the percolation of polyatomic species with the presence of impurities on a square lattice with periodic boundary conditions. Both bond and site percolation problems were taken into consideration. Linear $k$-mers, as well as so-called SAW $k$-mers, i.e. segments of a self-avoiding walk, have been studied up to values of $k=9$. A phase diagram where the critical concentration of impurities is plotted as a function of $k$ has been proposed. The concentration of impurities at which percolation becomes impossible, even at jamming coverage, decreases rapidly with increasing values of $k$. This research suggests that there is a critical length, $k$, at which percolation is impossible even without impurities. This suggestion may be incorrect if the critical concentration of impurities decreases to zero asymptotically  with increasing $k$. In principle, an investigation of the critical impurity concentration for large values of $k$ might answer the question: `Is the percolation of long rods possible?'.

The goal of the present research is to investigate the effect of different sorts of defects on percolation and jamming in the random sequential adsorption of linear $k$-mers onto a square lattice.
We analyzed and compared two different models. We investigated the RSA of (a) perfect linear $k$-mers onto a square lattice with previously placed non-conducting point defects ($L_d$ model), and (b)  linear $k$-mers with defects onto an ideal square lattice ($K_d$). The length of the $k$-mers, $k$, varied from 2 to 256, and periodic boundary conditions were applied.
For each given value of $k$, we were looking for a critical concentration of defects, $d_m$, when percolation is possible only at the jamming concentration.

The main question to be answered for both systems, i.e., the ideal $k$-mers on a lattice with defects and for  $k$-mers with defects on an ideal lattice, is the following: how do the percolation threshold and the jamming concentration vary with the length of the $k$-mers and the concentration of defects?

The rest of the paper is constructed as follows. In Section~\ref{sec:methods} we describe the technical details of our simulations. Section~\ref{sec:results} presents our main findings. In Section~\ref{sec:conclusion}, we summarize the results and conclude the paper.

\section{Details of the simulation\label{sec:methods}}
We considered a discrete two-dimensional space (square lattice $L \times L$) with periodic boundary conditions, i.e. we considered the RSA on a torus. The deposited objects were linear $k$-mers (particles occupying $k$ adjacent sites).
An object can be deposited in two allowed perpendicular orientations (vertical or horizontal) with equal probabilities, i.e. we are considering isotropic deposition. The filling fraction of the lattice by $k$-mers is $f= Nk/L^2$, where $N$ is the number of $k$-mers.
Two different models with non-conducting point defects on the lattice ($L_d$ model) and defects of the $k$-mers ($K_d$) are analyzed.

\subsection{Ideal $k$-mers on a lattice with defects, $L_d$ model}
The lattice is assumed to be initially filled with point nonconducting defects at a given concentration $d$.
Then the ideally conducting $k$-mers are deposited onto the substrate using RSA rules. We will denote this model as the $L_d$ model for short. The filling fraction of the lattice, $f$, changes within $[0;p_j]$, where $p_j(d)$ is the jamming concentration at the given $d$. In this model all $k$-mers are conducting and the fraction of conducting sites, $p$, coincides with $f$, i.e., $p=f$.

The jamming state for the $L_d$ model is presented in Figure~\ref{fig:lattices}b.

\subsection{$k$-mers with defects on an ideal lattice, $K_d$ model}
In this model, ideal lattice is filled using $k$-mers with defects. We will denote this model as the $K_d$ model for short.
The $k$-mers with defects contain some fraction of non-conducting defects. Let the average fraction of defects per object be $d$. The filling fraction of the lattice, $f$ changes within $[0;p_j]$ where $p_j$ is the jamming concentration for $k$-mers deposited on the ideal lattice.
The jamming state for the $K_d$ model is presented in Figure~\ref{fig:lattices}c.
In this model the fraction of conducting sites is $p = f ( 1 - d )$ and the connectivity analysis is carried out by accounting only for the conducting particles. The percolation concentration is determined as $p_c = f_c ( 1 - d ) $, where $f_c$ is the critical filling fraction that corresponds to the formation of a  spanning cluster of sites filled with conducting particles. At the jamming concentration, i.e., at $f= p_j$, percolation is observed when the concentration of defects is smaller than some maximum value, $d_m$. The percolation concentration for this case is  $p_c^m = p_j ( 1 - d_m ) $.

\subsection{Common technical details}

\begin{figure}%[htbp]
  \centering
  (a)\includegraphics[width=0.9\linewidth]{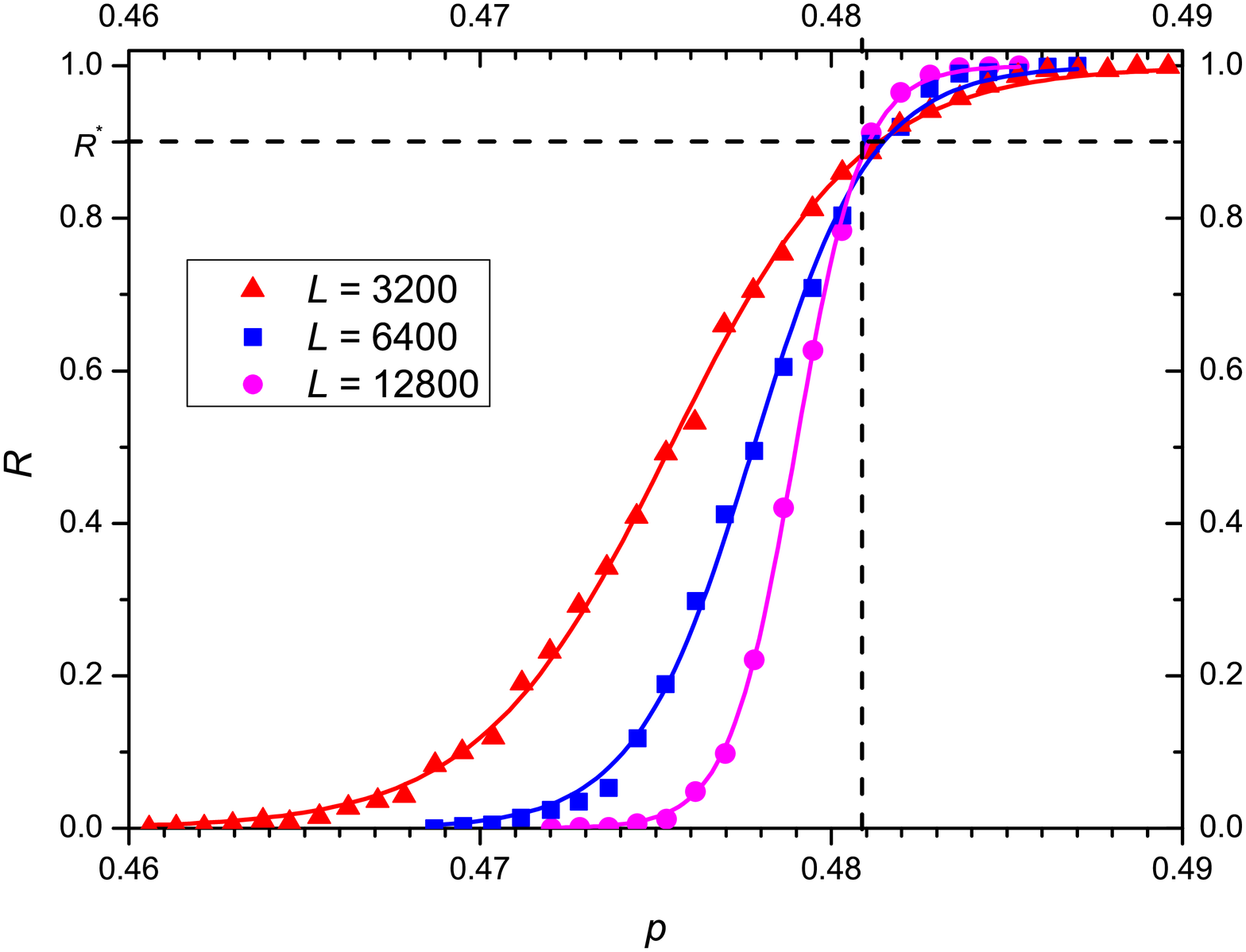} \\
  \hfill(b)\includegraphics[width=0.9\linewidth]{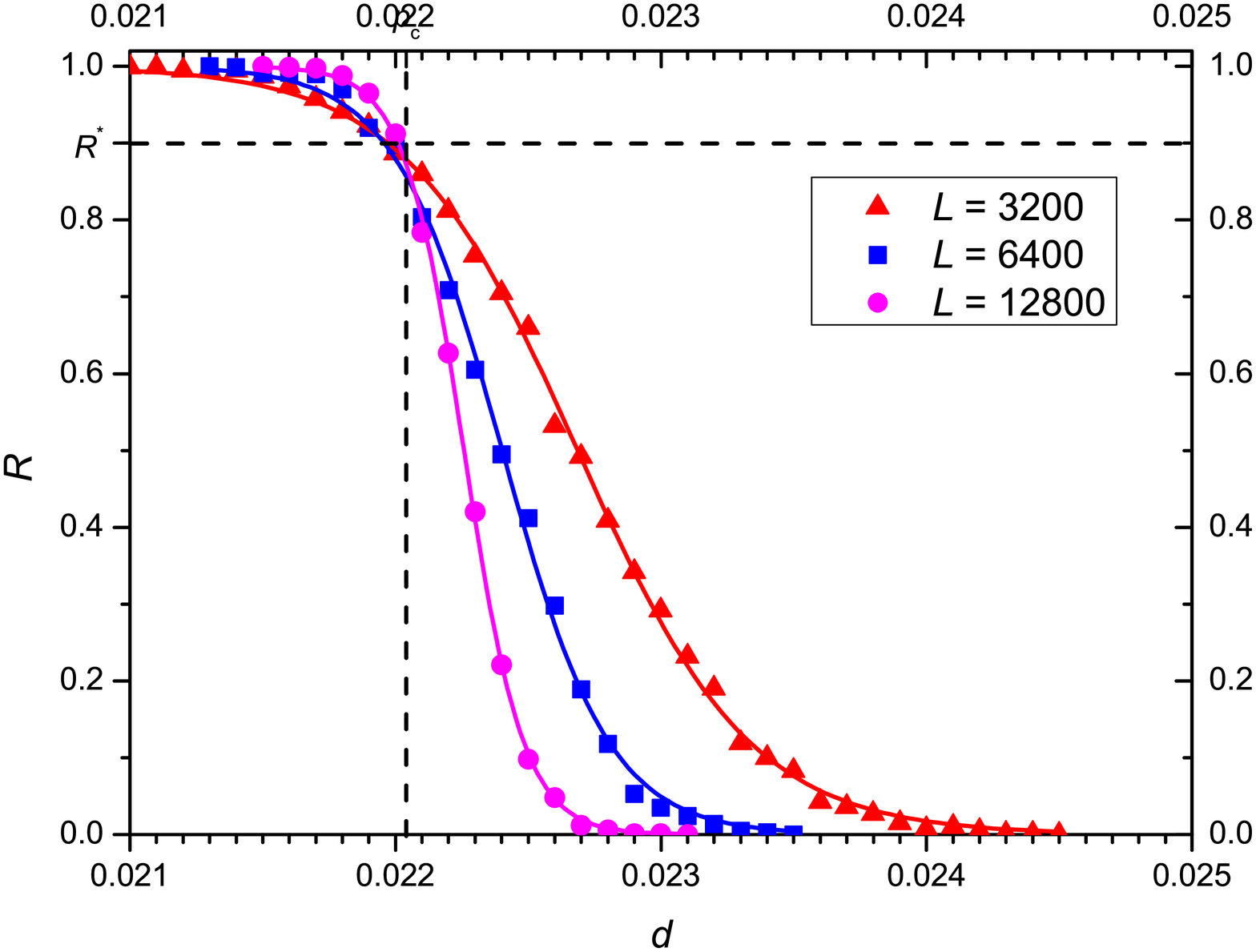}\\
  \caption{Examples of probability curves for the $L_d$ model for the concentration of objects (a) and defects (b). $k=32$. Criterion AND is used. The statistical error is smaller than the marker size. The solid lines correspond to fitting functions~\eqref{eq:sigmoidp} and \eqref{eq:sigmoidd}. The vertical dashed lines correspond to $p_c$ (a) and $d_m$ (b). The horizontal dashed lines correspond to $R(p_c)=R^*$ (a) and $R(d_m)=R^*$.\label{fig:fit}}
\end{figure}

The lattice is filled to a given concentration and then checked for the presence of percolation cluster or not.
Spiral clusters are treated as being percolating (see details in~\cite{Tarasevich2012PRE}).
In our simulation, a square lattice of size $L \times L$ sites was filled with $k$-mers until jamming coverage occurred. This was repeated 1000 times and the probability, $R_L$, for a cluster of k-mers wrapping the system was found. We used two criteria: criterion OR means there is a cluster wrapping the system either in a horizontal or vertical direction, criterion AND means there is a cluster wrapping the system in both directions simultaneously. The probability curves $R_L(p)$ and $R_L(d)$ have been fitted by the Boltzmann function
\begin{equation}\label{eq:sigmoidp}
R_L(p) = 1 - \left(1 + \exp\left(\frac{p - p_c(L)}{\Delta p}\right)\right)^{-1},
\end{equation}
\begin{equation}\label{eq:sigmoidd}
R_L(d) = \left(1 + \exp\left(\frac{d - d_c(L)}{\Delta d}\right)\right)^{-1},
\end{equation}
where $p_c(L)$, $d_c(L)$, $\Delta p$, and $\Delta d$ are the fitting parameters.
For examples of the analysis for the  $L_d$ model see Figure~\ref{fig:fit}.

We used three different lattice sizes to perform a scaling analysis and to find the percolation threshold in the thermodynamic limit ($L \to \infty$)(see, e.g.~\cite{Stauffer})
$$
p(L) -p_c(\infty))\sim L^{-1/\nu},
$$
where $\nu$ is the universal critical exponent. For percolation in two dimensions $\nu = 4/3$ (see, e.g.~\cite{Stauffer}). A similar relation was used for  assessing the critical concentration of defects in the thermodynamic limit.

\begin{figure}%[htbp]
  \centering
  a)\includegraphics[width=0.95\linewidth]{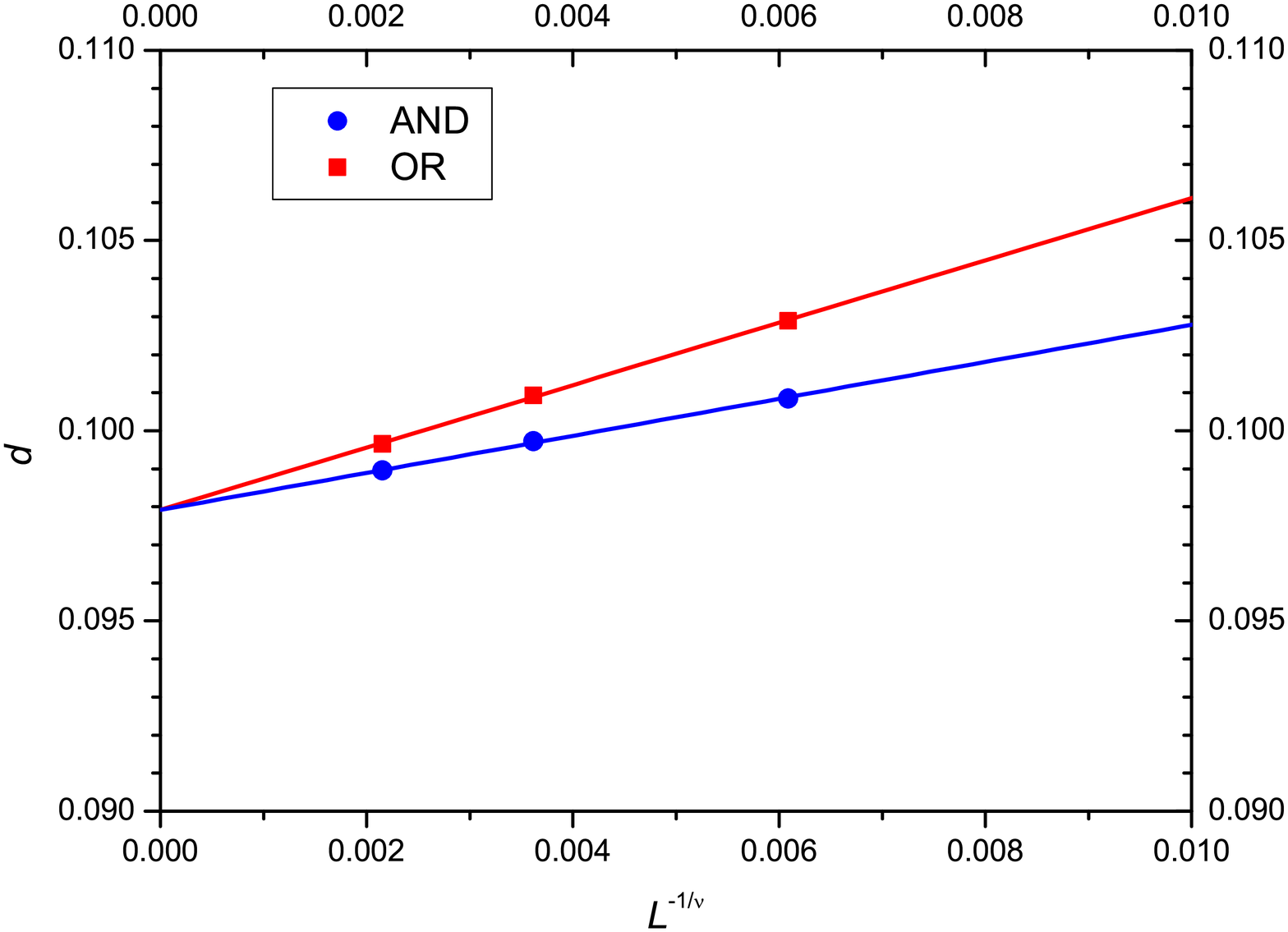}\\
  b)\includegraphics[width=0.95\linewidth]{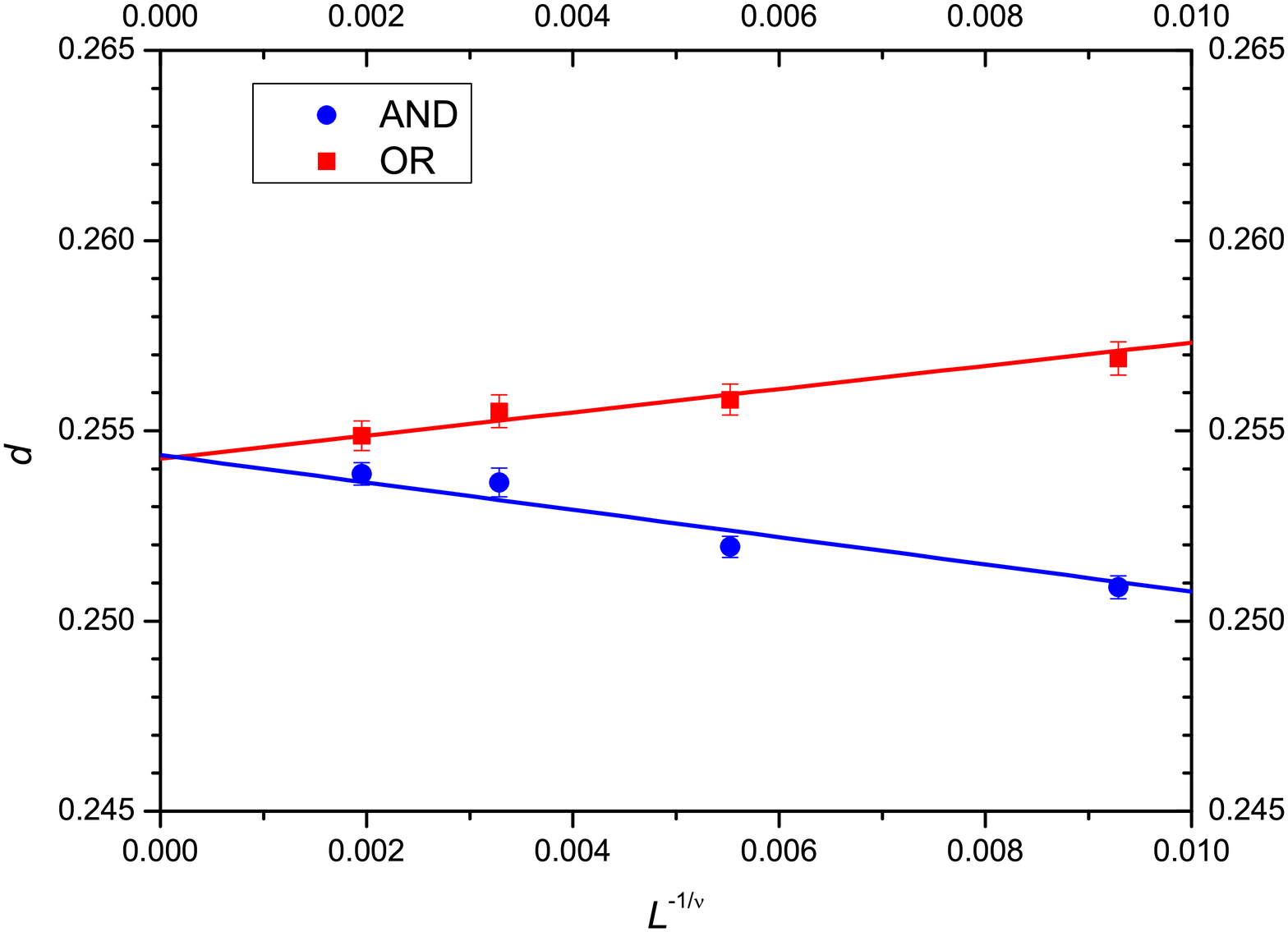}\\
  \caption{A sample of scaling  ($k = 9$) for the criterion AND (there is a wrapping cluster in both directions) (circles) and OR (there is a wrapping cluster either in the vertical or in the horizontal direction) (squares). $d$ is the concentration of defects, $L$ is the lattice size, $\nu$ is the critical exponent. (a) $L_d$ model. The statistical error is smaller than the marker size. (b) $K_d$ model.\label{fig:k9ORAND}}
\end{figure}

Examples of scaling for the criterion AND (there is a wrapping cluster in both directions), and OR (there is a wrapping cluster either in the vertical or in the horizontal direction) are shown in Figures~\ref{fig:k9ORAND}a ($L_d$ model) and~\ref{fig:k9ORAND}b ($K_d$ model).

For $k \leq 32$, we used lattice sizes $L=100k, 200k,$ and $400k$ to perform the scaling analysis. We obtained $p_c(L)$ and $d_c(L)$ using the probability curves for the criterion AND. For $k=64$, we used lattice sizes $L=100k, 150k$, and $200k$  to save time. To ensure precision, we utilized the probabilities for both the criteria AND and OR for scaling.
For $k=128$ the lattice sizes $L=75k, 100k$, and $125k$ and the probabilities for both the criteria AND and OR were used. Additionally, the critical concentrations of defects and objects were estimated using $R^* \approx 0.9$ ($R_L(p_c) = R^*$)~\cite{Tarasevich2012PRE}.

\begin{figure}%[htbp]
  \centering
  \includegraphics[width=0.9\linewidth]{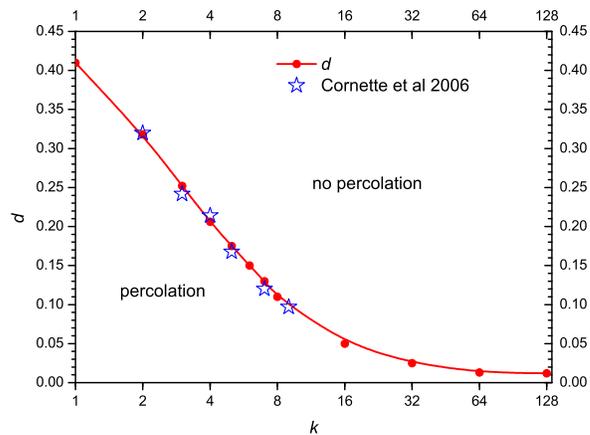}\\
  \caption{Comparison of our results (circles) and the results obtained earlier~\cite{Cornette2006JCP,Cornette2006PLA} (stars). $d_m$ is the critical concentration of defects  which blocks percolation even at jamming concentration.  The error bars are smaller than the symbol size. The curve corresponds to our fitting function~\eqref{eq:fit} ($L_d$ model). The statistical error is smaller than the marker size. \label{fig:Cornette}}
\end{figure}

To verify our program and method, we compared our results for the isotropic deposition with published numerical simulation results for short linear $k$-mers ($k \leq 9$) on a diluted square lattice ($L_d$ model)~\cite{Cornette2006PLA,Cornette2006JCP} (Figure~\ref{fig:Cornette})).
The agreement of our results and the published data is very good.

\section{Results and Discussion\label{sec:results}}
\begin{figure}%[htbp]
  \centering
  \includegraphics[width=0.9\linewidth]{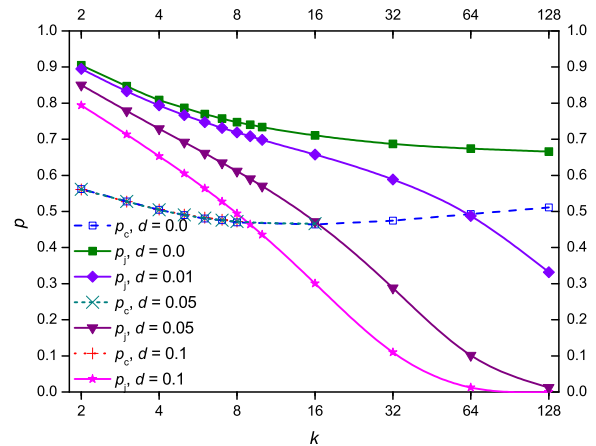}
  \caption{Percolation threshold and jamming concentration vs length of $k$-mers for different defect concentrations. $L_d$ model. The lines are a guide to the eye.\label{fig:pcpsvsk}}
\end{figure}

 Figure~\ref{fig:pcpsvsk} demonstrates how the percolation threshold and jamming concentration vary with the length of the perfect $k$-mers in the presence of defects on the substrate at a given concentration. The jamming concentration of $k$-mers decreases when the concentration of defects grows. The larger the value of $k$ the more noticeable is the effect.  By contrast, the percolation threshold is almost insensitive to the defect concentration.

\begin{figure}%[htbp]
  \centering
  (a)\includegraphics[width=0.95\linewidth]{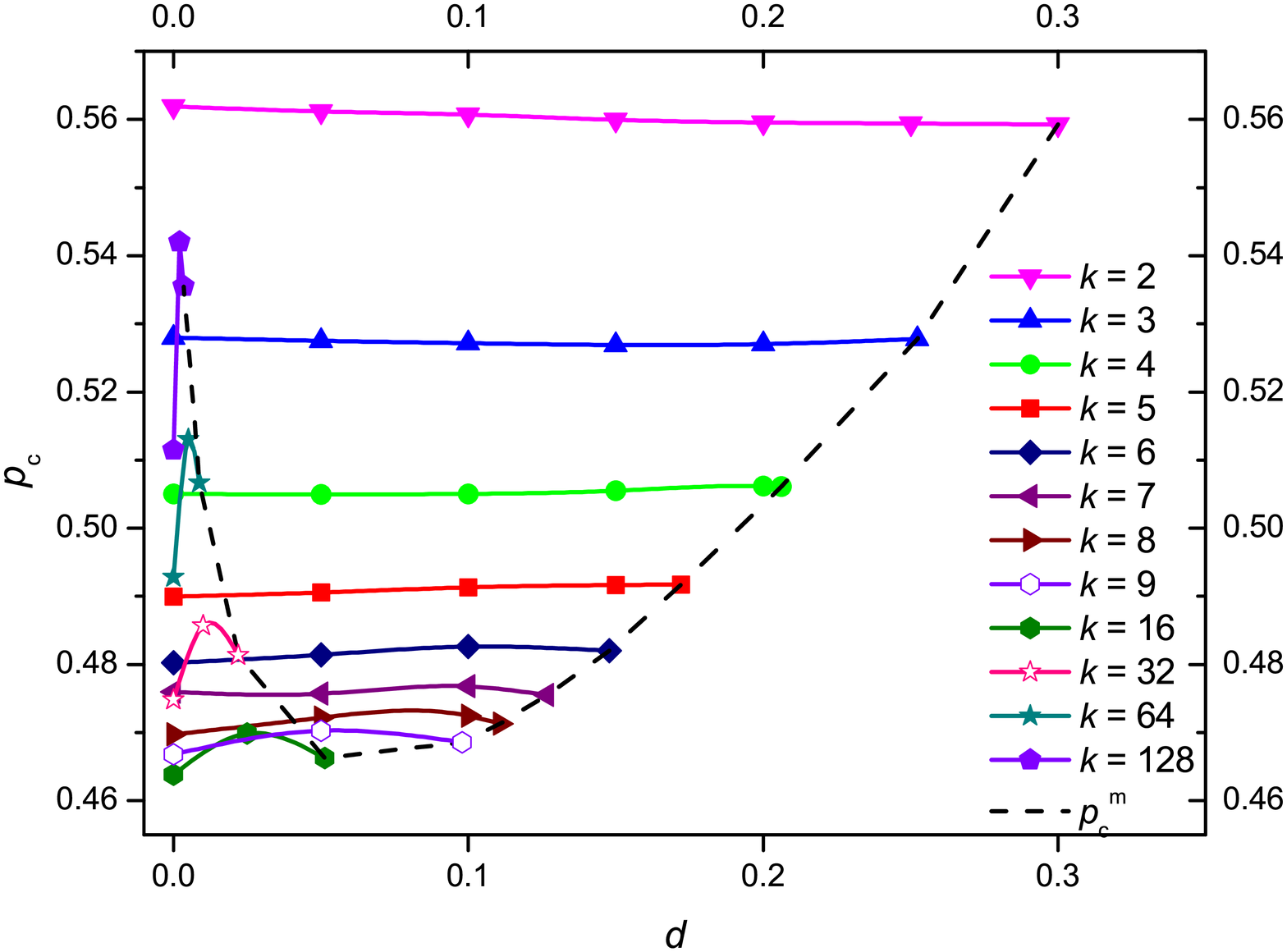}\\
  (b)\includegraphics[width=0.95\linewidth]{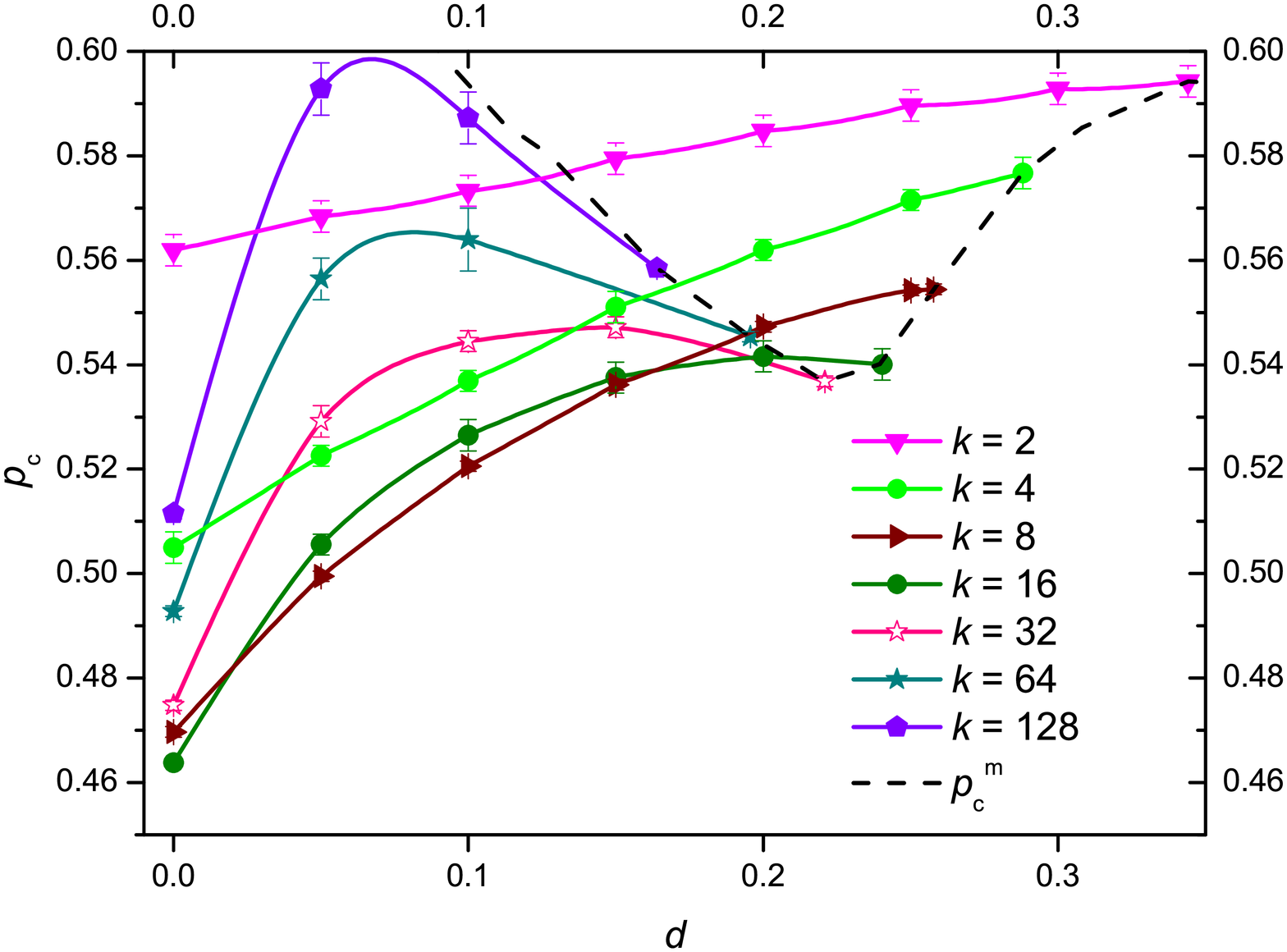}\\
  \caption{Percolation threshold $p_c$ vs concentration of defects, $d$, for different values of $k$. The solid lines are a guide to the eye. The dashed lines show dependencies $p_c(d_m)$, where $d_m$ is a critical concentration of defects. (a) $L_d$ model. The statistical error is smaller than the marker size. (b) $K_d$ model.
  \label{fig:pcvsd}}
\end{figure}

Figure~\ref{fig:pcvsd} presents the percolation threshold $p_c$ vs the concentration of defects $d$ for different values of $k$. Percolation is observed when the concentration of defects is smaller than a  critical concentration of defects, $d_m$. In the $L_d$ model, the percolation threshold is almost independent of the defect concentration up to $k \approx 10$ and has a maximum ($d \in [0,d_m]$) for larger values of $k$ (Figure~\ref{fig:pcvsd}a). In contrast, the percolation threshold in the $K_d$ model is rather sensitive to the defect concentration. It increases monotonically between $d=0$ and $d=d_m$ for short $k$-mers ($k<8$) but has a distinct maximum for longer objects ($k>8$) (Figure~\ref{fig:pcvsd}b). The observed differences evidently reflect distinctions in the configurations of the $k$-mers for the two models studied. E.g., in the jammed state, the two models $L_d$ and $K_d$ produce rather different configurations of deposited $k$-mers. In the $L_d$ model, the defects hinder the further deposition of the $k$-mers and the jammed state for this model (Figure~\ref{fig:lattices}a) is more spars compared with the conventional jamming that is realized with the $K_d$ model (Figure~\ref{fig:lattices}b).

\begin{figure}%[htbp]
  \centering
  \includegraphics[width=0.9\linewidth]{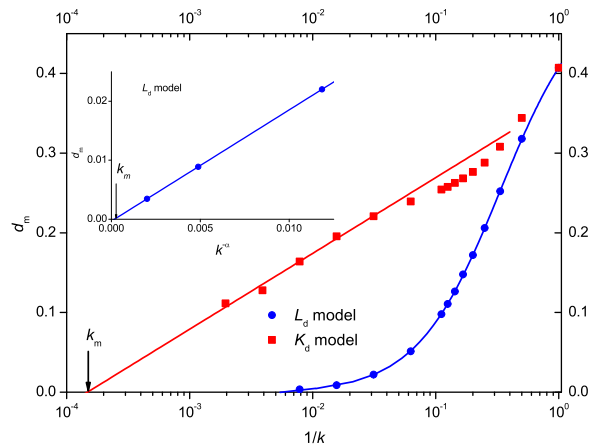}\\
  \caption{Critical concentration of defects vs reciprocal length of $k$-mers. $L_d$ and $K_d$  models. The statistical error is smaller than the marker size. For the $L_d$ model, the solid line corresponds to the fitting function~\eqref{fig:fit}. For the $K_d$ model, the solid line corresponds to the linear fitting functions. Inset:  Magnification of long $k$ region for the $L_d$ model. \label{fig:d1k}}
\end{figure}

A graphical representation (Figure~\ref{fig:d1k}) is more suitable for the analysis.

For the $L_d$ model, the data are well fitted with a high coefficient of determination ($R^2 = 0.9999$)
by the function
\begin{equation}\label{eq:fit}
d_m = a \frac{ k_m^{\alpha} - k^{\alpha}}{b  + k^{\alpha}}
\end{equation}
with $a = 0.006 \pm 0.002$,
$\alpha = 1.28 \pm 0.01$,
%$B = 1.968 \pm 0.007$, and $C = 0.486\pm 0.006$.
$b = 3.75 \pm 0.14$. %$k_m = 5900 \pm 500$ %$A = (-7 \pm 4) \times 10^{-4}$,
The critical length of the$k$-mers, $k_m $, is approximately $ 5900 \pm 500$.

For long $k$-mers in the limit of $k\to k_m$, Eq. \ref{eq:fit} gives $d_m \propto k^{-\alpha}$ (see, Figure~\ref{fig:d1k}b).

On the other hand, for the $K_d$ model, in the limit of large values of $k$
the data are well fitted ($R^2 = 0.9998$) with the following function
\begin{equation}\label{eq:fitKd}
d_m = a\log(k_m/k)
\end{equation}
with $a = 0.103 \pm 0.006$, $k_m = 4700 \pm 1000$ (see, Figure~\ref{fig:d1k}c).

Note that the estimated critical lengths of $k$-mers are in close correspondence with the earlier  estimations~\cite{Kondrat2001PRE,Tarasevich2012PRE}.

Table~\ref{tab:values} summarizes our results on the critical concentration of defects ($d_m$) at which percolation is possible only at jamming concentration ($p_c^m$) for the $L_d$ and $K_d$ models. The percolation thresholds for the ideal $k$-mers on an ideal square lattice are also presented for comparison.

\begin{table}%[htbp]
  \centering
  \caption{Critical concentration of defects ($d_m$) at which percolation is possible only at the jamming concentration ($p_c^m$) and corresponding percolation threshold ($p_c(d=0)$) for linear $k$-mers. $L_d$ and $K_d$ models.\label{tab:values}}
  \begin{ruledtabular}
\begin{tabular}{rllllll}
&&&\multicolumn{2}{c}{$L_d$ model}&\multicolumn{2}{c}{$K_d$ model}\\
$k$ & $p_c(d=0)$& $p_j(d=0)$ &$p_c^m$ & $d_m$ &$p_c^m$ &  $d_m$ \\
\hline
  1 & 0.5927  & 1      & 0.5927  & 0.4073  & 0.5927 & 0.4073\\
  2 & 0.5619  & 0.906  & 0.55930 & 0.31803 & 0.5619 & 0.3441\\
  3 & 0.52797 & 0.846  & 0.52780 & 0.252200& 0.538  & 0.308\\
  4 & 0.5050  & 0.81   & 0.50614 & 0.20612 & 0.522  & 0.288\\
  5 & 0.48997 & 0.7868 & 0.49176 & 0.17219 & 0.5103 & 0.2765\\
  6 & 0.48026 & 0.7703 & 0.48205 & 0.14780 & 0.502  & 0.2683\\
  7 & 0.47600 & 0.7579 & 0.47552 & 0.12656 & 0.4953 & 0.2626\\
  8 & 0.4697  & 0.747  & 0.4713  & 0.11072 & 0.4892 & 0.2578\\
  9 & 0.4668  & 0.7405 & 0.46855 & 0.09792 & 0.4861 & 0.2544\\
 16 & 0.4638  & 0.71   & 0.46627 & 0.05136 & 0.4706 & 0.2394\\
 32 & 0.4748  & 0.689  & 0.48094 & 0.02202 & 0.4681 & 0.2209\\
 64 & 0.4928  & 0.678  & 0.50665 & 0.0089  & 0.4824 & 0.1956\\
128 & 0.5115  & 0.668  & 0.5355  & 0.0035  & 0.504  & 0.164\\
256 & 0.53    & 0.665  &         &         & 0.537  & 0.128\\
\end{tabular}
\end{ruledtabular}
\end{table}

%\clearpage

\section{Conclusion\label{sec:conclusion}}

It is well known that extrapolation is less reliable than interpolation. Conjectures for the percolation threshold and jamming concentration tend towards constants as the length of the $k$-mers increases, based on the simulations for relatively short $k$-mers~\cite{Becklehimer1992,Bonnier1994,Vandewalle2000epjb,Cornette2003epjb,Longone2012PRE}. However, this conjecture seems to become invalid  when considering the results of computer-based experiments with larger k-mers~\cite{Leroyer1994PRB,Kondrat2001PRE,Tarasevich2012PRE}. In any case, the percolation and jamming behavior of adsorbed layers produced by the deposition of very long $k$-mers ($k \gtrsim 10^3$) is  still not particularly clear because the conclusions are based on results obtained for not very large objects then being extrapolated to represent very long objects. Direct simulation of the RSA for such large objects looks like a very time-consuming task and, to the best of our knowledge, has not yet been undertaken.

The results presented in this work give an indirect confirmation of the conjecture that percolation of $k$-mers is impossible if the length exceeds a critical value. Our new estimation gives that critical length is a little bit less than the value published in earlier works~\cite{Kondrat2001PRE,Tarasevich2012PRE}.

From a practical point of view, further consideration of the RSA of objects with defects onto a substrate with defects  looks very attractive  as a research project. Moreover, the deposition of the objects onto a substrate may, in fact, actually be not isotropic.  We will report our results on this  more realistic model in future papers.

\section*{Acknowledgements}

The reported research is supported by the Ministry of Education and Science of the Russian Federation, project no~266/643, Russian Foundation for Basic Research, grant no 14-02-90402 Ukr\_a, and the National Academy of Sciences of Ukraine, project no 43–02–14(U).

\bibliography{percolation,RSA}

\end{document}